\documentclass{optica-article}

\journal{opticajournal} 

\articletype{Research Article}

\usepackage{graphicx} 
\usepackage{physics}
\usepackage{mathtools}
\usepackage{nicefrac}
\usepackage{siunitx} 
\usepackage{color, colortbl} 
\usepackage{xcolor, colortbl} 
\usepackage{wasysym}
\usepackage{multirow}
\usepackage[normalem]{ulem} 
\usepackage{float}
\usepackage{xr}
\usepackage{array}
\usepackage{hyperref}
\usepackage[T1]{fontenc}
\usepackage{notes2bib}
\usepackage{anyfontsize} 

\newcommand{\kk}{{\vb k}}

\newcommand{\pcm}[1]{\SI{#1}{\per\cm}}
\newcommand{\um}[1]{\SI{#1}{\micro\meter}}
\newcommand{\nm}[1]{\SI{#1}{\nano\meter}}
\newcommand{\pum}[1]{\SI{#1}{\per\micro\meter}}
\newcommand{\pct}[1]{\SI{#1}{\percent}}
\usepackage{xurl}

\begin{document}
\title{Momentum-resolved reflectivity of a 2D photonic crystal in the near-infrared} 

\author{Timon J. Vreman, Melissa J. Goodwin, Ad Lagendijk, Willem L. Vos\authormark{*}}
\address{Complex Photonic Systems (COPS), Faculty of S\&T, University of Twente, 7500 AE Enschede, The Netherlands}
\homepage{https://nano-cops.com}
\email{\authormark{*}w.l.vos@utwente.nl}

\begin{abstract*}
Two-dimensional (2D) photonic crystals offer strong control over the propagation of light through their bands.
Theoretical methods for computing the band structure in 2D are well-established and fast because 2D photonic crystals are homogeneous in the third dimension.
Experimental verification is scarce, however, especially in the telecom range, because real photonic crystals and experimental methods inherently cannot be homogeneous in the third dimension. 
In this work, we report momentum-resolved reflectivity measurements on photonic crystals that are periodic in two dimensions and homogeneous over a thickness of \SI{5}{\micro\meter}.
Using Fourier spectroscopy, we carefully select wave vectors in the 2D plane of periodicity of the photonic crystal.
Our experiments agree excellently with 2D band structure calculations and with 2D finite-difference time-domain simulations, confirming that our experimental methods truly pertain to nanophotonics in 2D.
Our results provide a robust bridge between theory and experiment, and our techniques can be readily extended to other 2D structures, including those with functional defects.
\end{abstract*}

\section{Introduction}
Photonic crystals are complex nanophotonic structures with a periodic refractive index $n(\vb{r})$ that enable control over the propagation and emission of light~\cite{Joannopoulos2008book, Saleh2019Wiley}.
Due to their periodicity on the scale of the wavelength of light, photonic crystals exhibit Bloch modes with unique optical properties, such as slow light or the opportunity for photonic gaps~\cite{yablonovitch1993JOSAB, baba2008NatPhot, krauss2007JPDAP, li2008OpEx}; properties that are often complex to study.
Whereas 3D photonic crystals have a periodic refractive index in three dimensions, 2D photonic crystals are periodic in two dimensions $(x,y)$ and homogeneous in the third $(z)$.
Still, 2D photonic crystals pertain to optical properties of photonic crystals like bands, band gaps, and slow light.
In addition, the polarizations of light in 2D structures can be separated as opposed to 3D, making the complex properties of photonic crystals easier to understand.
Therefore, photonic crystals are often studied theoretically in 2D~\cite{Sakoda2005Springer, Notomi2000prb, foteinopoulou2005PRB}.

Experimentally probing the modes of 2D photonic crystals to compare with 2D theory is remarkably complex because the third dimension of the samples should be effectively homogeneous.
One approach is to use long wavelengths and a large lattice parameter, making the manufacturing of structures that are homogeneous over a large $\Delta z$ of \um{100} or more possible~\cite{Robertson1992PRL, leonard2000prb, galli2002PRB, chalimah2021PRAp}.
However, optical techniques are challenging when using the required long wavelengths, and the accompanying large focus spot readily obscures interesting local features such as intentional and unintentional lattice defects. 
Furthermore, a more accessible approach that functions in the telecom range would strongly support integration with present and anticipated uses of photonic crystals.

\begin{figure}[t!]
    \centering
    \includegraphics[width=0.6\linewidth]{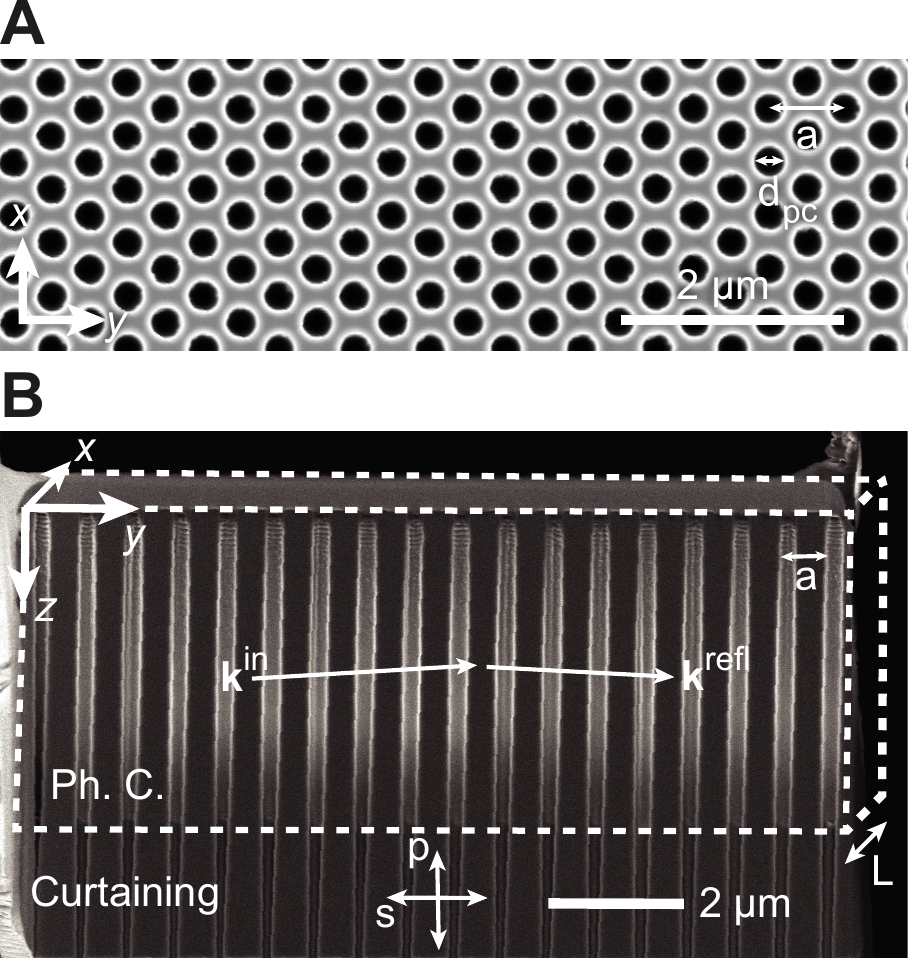}
    \caption{Scanning electron microscope (SEM) images of (a) our bulk 2D photonic crystal viewed from the top and (b) our thin slice viewed from the front.
    (White dashed lines) Box indicating the location of the 2D photonic crystal.
    Vectors $\kk{}^{\rm{in}}$ and $\kk{}^{\rm{refl}}$ are examples of a possible in-plane incoming angle and reflected angle.
    The $s$- and $p$-polarized directions are defined at the bottom.
    Due to the focused ion beam cutting technique, the pores at the surface extend into `curtains'.
    }
    \label{fig:sem_T2_D}
\end{figure}

Fourier imaging is a tool for resolving the wave vector $\vb{k}$ dependence of optical reflection~\cite{kurvits2015JOptSocAmA, vasista2019book, Zhang2021ScienceB, cueff2024Nanophotonics}.
Contrary to standard angle-resolved reflectivity measurements, many wave vectors $\vb{k}$ from the high numerical aperture (NA) objective are examined simultaneously.
The focus spot is small so the reflectivity is only probed locally.
Fourier imaging and standard angle-resolved imaging have been applied to measure the reflectivity of 2D photonic slab waveguides, probing the bands from the $z$ direction~\cite{astratov1998IEEproc, galli2004prb, le_thomas2007JOptSocAmB, zhang2018PRL}.
However, slab waveguides are not captured by 2D theory because they are not homogeneous in $z$~\cite{notomi2002IEEE, li2008OpEx, hou2011IEEE}.
In addition, the 2D in-plane modes with $k_z = 0$ cannot be probed from the $z$ direction.
Moreover, mode excitation effects at a 2D air-crystal interface cannot be studied either in this configuration, while many complex effects involving, for example, symmetry and group velocity often occur at this interface~\cite{Notomi2000prb, Sakoda2005Springer, Joannopoulos2008book, foteinopoulou2005PRB}.

Therefore, we here present momentum-resolved reflectivity measurements of 2D photonic crystals, periodic in $x$ and $y$, and homogeneous for a range $\Delta z =$ \um{5}, see Fig.~\ref{fig:sem_T2_D}.
We probe the crystals with a small focus spot using Fourier spectroscopy, such that our thin 2D photonic crystal is described by purely 2D theory.
Our experimental dataset contains a wealth of information, clearly displaying the dispersion of 2D Bloch modes in the reflectivity spectra.
This information can be used to further understand the properties of real photonic crystals and the intricacies at air-crystal interfaces.
Furthermore, we validate our experiments with 2D momentum-resolved finite-difference time-domain simulations and photonic band calculations with symmetry analysis.
Our results establish Fourier spectroscopy as an accurate and efficient tool for studying 2D photonic systems, excellent for uniting theory and experiment.

\section{Methods}
\subsection{Samples}
Our periodic 2D photonic crystals shown in Fig.~\ref{fig:sem_T2_D} consist of air pores in silicon.
The nanostructures are created from high-purity silicon wafers with a chromium etch mask with a centered rectangular lattice with lattice parameters ($a, c$) where $a$ = \nm{680} and $c = a/\sqrt{2}$, which were previously prepared in our group~\cite{broekvanden2012AdvFunctMater}.
The samples are etched using deep reactive-ion etching (SPTS Pegasus) using the method detailed by Goodwin \textit{et al.}~\cite{goodwin2023Nanotech}.
After etching, the pores are $L =$ \um{5} deep, have a diameter $d$ = \nm{244\pm20}.
Importantly, each pore is close to identical and almost perfectly cylindrical.

For transmission experiments, it is necessary to have a sample that is sufficiently thin in the $x$ direction for a significant fraction of the light to be transmitted.
To this end, a Focused Ion Beam (FIB) was used to lift out a slab of 2D photonic crystal.
A protective layer of carbon was applied to the slab to prevent surface damage from the ion beam, and the slab was welded to a copper lift-out grid (Omniprobe) using carbon.
The slab was thinned front and back at \SI{30}{\kilo\volt} and \SI{90}{\pico\ampere}, resulting in a slab-shaped sample with a thickness of $L$ = \um{2.65\pm0.35} $= 5.5 c$.
Due to the FIB milling, the pores extend into curtains (see Fig.~\ref{fig:sem_T2_D}), which are located below the photonic crystal and are only present at the surfaces.
Experiments are repeated using a different photonic crystal from another wafer, yielding similar results.

\subsection{Optical methods}
\begin{figure}[t]
    \centering
    \includegraphics[width=0.45\linewidth]{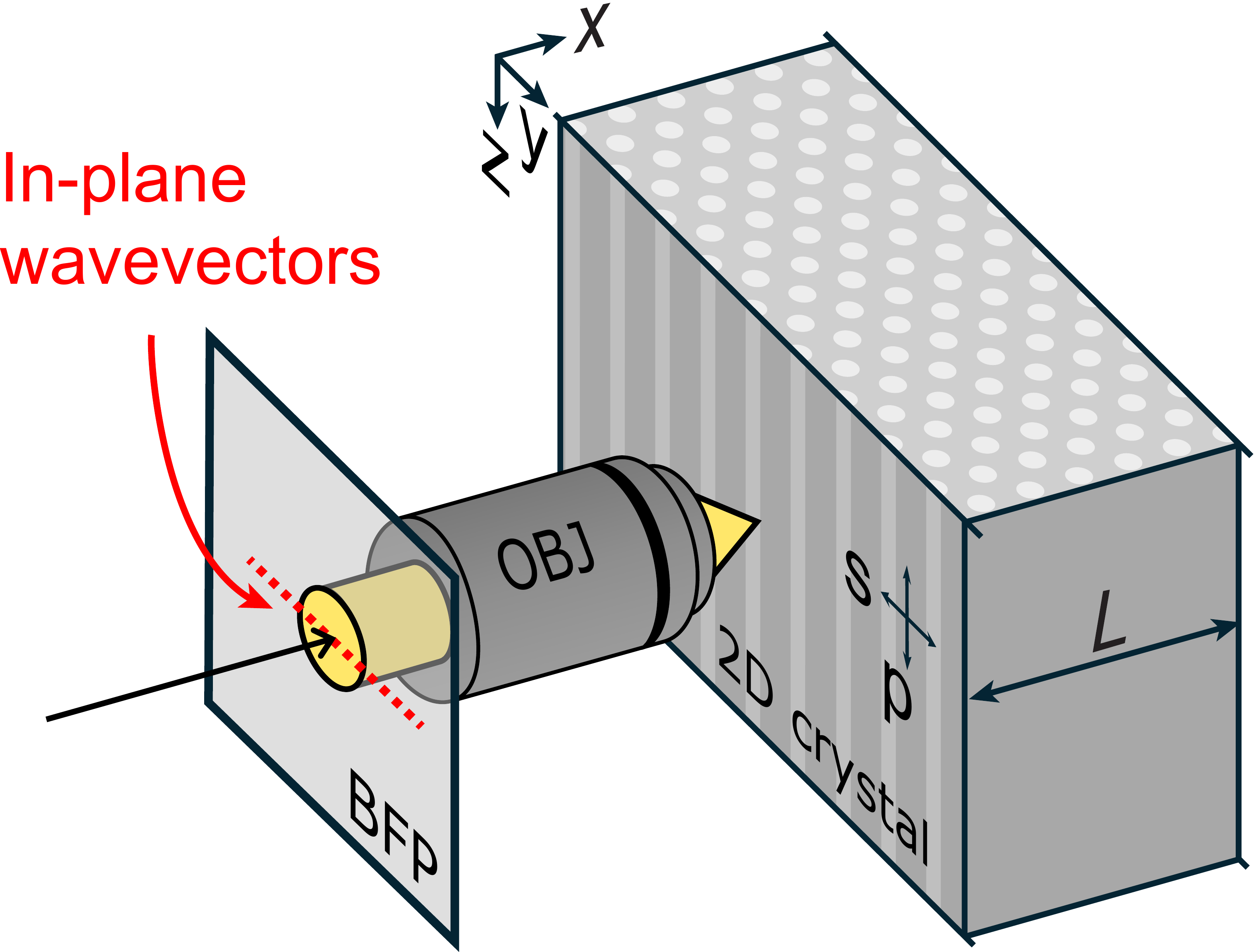}
    \caption{Schematic of the setup.
    A plane wave is incident on the back focal plane (BFP) of the objective (OBJ), which focuses the light onto the 2D photonic crystal.
    The objective also collects reflected light. 
    The back focal plane is imaged onto a camera, of which only light in the plane of the 2D sample is selected (red dashed line).
    The $s$- and p-polarizations are defined at the right.
    The illustration is not to scale.
    }
    \label{fig:pc_optSetup}
\end{figure}

{\label{sec:qcry2D_opticalMethods}}
We use the momentum-resolved setup shown in Fig.~\ref{fig:pc_optSetup}, which employs Fourier spectroscopy~\cite{cueff2024Nanophotonics, zhang2022prb}.
Plane waves with tunable wavelengths $\lambda$ between 900 and \nm{1650} and a bandwidth $\Delta\lambda=$ \nm{0.6} are incident on the back focal plane of a 100$\cross$ objective with a numerical aperture NA = 0.85.
A half-wave plate sets the polarization to $s$- or $p$-polarized light, where the electric field points in the y- or z-direction, respectively.
The objective focuses the light to a spot with a diameter FWHM (full width at half maximum) of $d_f$ $<$~\um{2} and picks up the reflected light in backscattering geometry.
The back focal plane is imaged onto an InGaAs camera (Photonic Sciences).
From our optical measurements, we obtain a 3D data set consisting of intensity as a function of the wavenumber $\tilde{\nu}$ ($\equiv 1/\lambda$), and wave vector components $k_y$ and $k_z$, see Fig.~\ref{fig:pc_optSetup}.
To calibrate the reflectivity in percentage points, the intensity collected from the sample is divided by the intensity reflected on a gold mirror assuming a reflectivity of $R_{\rm{Au}} =$ \pct{96}.

We manage to measure reflectivity in the 2D ($x$,$y$) plane of periodicity as follows:
Incident wave vectors with $k^{\rm{in}}_z = 0$, i.e., in the 2D plane of periodicity, are reflected toward any $k^{\rm{refl}}_x$ and $k^{\rm{refl}}_y \equiv k_y$ depending on the 2D structure, but they remain in the 2D plane because $n(\vb{r})$ does not depend on $z$.
Likewise, incident out-of-plane wave vectors with $k^{\rm{in}}_z \neq 0$ are diffracted to outgoing out-of-plane wave vectors with $k^{\rm{refl}}_z = k^{\rm{in}}_z \neq 0$.
Therefore, from the 3D momentum-resolved reflectivity, we select wave vectors with $k^{\rm{refl}}_z = 0$ to obtain the 2D momentum-resolved reflectivity.
We probe the intensity at particular reflected wave vectors $\kk{}^{\rm{refl}} \equiv \kk{}$ relative to that on a gold mirror.
It is possible to obtain more than \pct{100} reflectivity at a particular $\kk{}$ if multiple $\kk{}^{\rm{in}}$ are diffracted toward the same $\kk{}$, i.e., when incident light is redirected.
However, the reflectivity at a single wavenumber integrated over all $\kk{}$ is bounded to \pct{100}.

\subsection{Band calculations}\label{sec:pc_bands}
For our calculations, we use $a =$ \nm{680}, $d$ = \nm{244}, $n_{\rm{Si}} = \sqrt{12.1}$, and the primitive unit vectors of the centered rectangular lattice.
We will first estimate which Bragg plane primarily determines the reflectivity in our experimental spectral detection regime centered around $\lambda_c =$ \nm{1300} (wavenumber $\tilde{\nu} =$ \pcm{7690}).
According to first-order Bragg reflection for normal incidence $\lambda = 2nd_B$, where $d_B$ is the spacing of lattice planes.
Given $\lambda_c =$ \nm{1300} and volumetric average refractive index $n_{\rm{avg}} =$ 2.77, we find $d_B =$ \nm{235} $\approx c/2 =$ \nm{240}.
Therefore, we estimate that we primarily probe the bands with $\vb{k}$-vectors on the $hk =$ 11 Bragg plane.
We calculate the bands along this Bragg plane by the open-source package MIT Photonic Bands (MPB)~\cite{johnson2001OptExp}.

\begin{figure}
    \centering
    \includegraphics[width=0.5\linewidth]{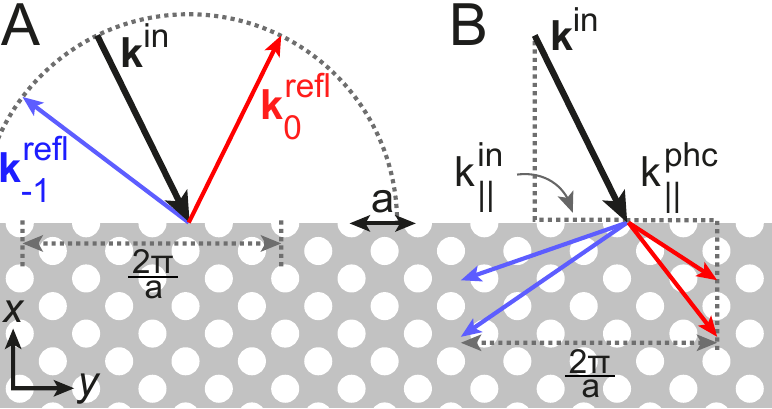}
    \caption{(a) Reflection and (b) refraction on a 2D photonic crystal.
    (a) Specular reflection $k^{\rm{refl}}_0$ (red) and an example of higher order diffraction $k^{\rm{refl}}_{-1}$ (blue) from $k^{\rm{in}}$ (black).
    (b) Examples of excited Bloch modes, where $k^{\rm{in}}_\parallel = k^{\rm{phc}}_\parallel + m\frac{2\pi}{a}$, where $m \in \mathcal{Z}$.
    Note that Bloch modes have periodic $\vb{k}^{\rm{phc}}$, so the blue and red modes are identical.
    }
    \label{fig:pc_reflectionExplained}
\end{figure}

Incident waves can excite Bloch modes.
If a Bloch mode is excited, the reflectivity is expected to be low, and conversely, if no Bloch mode is excited, the reflectivity is high.
Therefore, in our experiments, we infer the excitation of Bloch modes from the absence of reflected light.

Due to the periodic surface, diffraction orders are possible, where $k^{\rm{in}}_\parallel = k^{\rm{refl}}_{\parallel} + l\frac{2\pi}{a}$, where $l \in \mathcal{Z}$, see Fig.~\ref{fig:pc_reflectionExplained}(a).
To excite a Bloch mode, the incident wave must have the same wavenumber as the Bloch mode of the photonic crystal, $\tilde{\nu}^{\rm{in}} = \tilde{\nu}^{\rm{phc}}$, and a wave vector $k^{\rm{in}}_y = k^{\rm{phc}}_y + m\frac{2\pi}{a}$, where  where $m \in \mathcal{Z}$~\cite{Notomi2000prb}, illustrated in Fig.~\ref{fig:pc_reflectionExplained}(b).
In addition, the symmetry properties must match as discussed further~\cite{Sakoda2005Springer}, and the phase velocity of the Bloch modes must not point toward the surface~\cite{foteinopoulou2005PRB}.
The phase velocity is not considered here because it requires a detailed investigation using equifrequency surfaces; therefore, we may calculate a few additional modes that we cannot excite.
The amplitudes of reflected and excited waves are determined by the interface, and are computed later in Sec.~\ref{sec:pc_fdtd}.

Non-specularly reflected diffraction orders such as $\vb{k}^{\rm{refl}}_{-1}$ in Fig.~\ref{fig:pc_reflectionExplained}(a) may interfere with the specularly reflected order $\vb{k}^{\rm{refl}}_{0}$, which would complicate data analysis.
We calculate at which wavenumbers higher orders are collected by the objective to show that such interference only occurs in a small portion of our collected momentum-resolved reflectivity spectra:
collection of higher diffraction orders occurs soonest when $k^{\rm{in}}$ and $k^{\rm{refl}}_{-1}$ overlap and coincide with the maximum collected $k_y$ of the objective, i.e., when $k^{\rm{in}}_y = (\rm{NA})\it{k}_{\rm 0} \ge \frac{\pi}{a}$, which occurs at wavenumber $\tilde{\nu} \ge$ \pcm{8650}.
For normal modes ($k^{\rm{in}}_y = 0$), higher orders may interfere only when $\tilde{\nu} \ge$ \pcm{17300}, far outside our detection regime.  
Therefore, the calculated bands with wavenumbers $\tilde{\nu}_{\rm{calc}}$ and off-axis wave vector $k^{\rm{calc}}_y$ can be directly plotted onto the experimentally obtained wavenumber versus off-axis wave vector graphs.

As mentioned, for an incident wave to excite a Bloch mode (see Fig.~\ref{fig:pc_reflectionExplained}(b)), their symmetry properties must match~\cite{Sakoda2005Springer, Vreman2025}.
One symmetry property is polarization, which is separable for 2D structures.
The most important direction for reflectivity is the direction normal to the sample, $\vb{k} = k_x\hat{\vb{x}}$.
The Bloch modes with $\vb{k}^{\rm{phc}} = k^{\rm{phc}}_x\hat{\vb{x}}$ have other symmetry in addition to polarization, namely whether they are symmetric or antisymmetric for a mirror flip in the $y$ direction, $m_{01}$.
As $s$-polarized ($p$-polarized) plane waves are antisymmetric (symmetric) for $m_{01}$, they can only excite antisymmetric (symmetric) Bloch modes.
Although the $m_{01}$ mirror flip only holds exactly at $\vb{k}^{\rm{phc}} = k^{\rm{phc}}_x\hat{\vb{x}}$, the $m_{01}$ flip property is likely to remain dominant for off-axis modes with $\vb{k}^{\rm{phc}} \approx k^{\rm{phc}}_x\hat{\vb{x}}$.
Therefore, only those states with symmetry properties that are excited by plane waves at $\vb{k}^{\rm{in}} = k^{\rm{in}}_x\hat{\vb{x}}$ are shown.
These symmetry-selected bands are expected to describe the most prominent features of the measured momentum-resolved reflectivity.

\subsection{Finite-difference time-domain simulations}\label{sec:pc_fdtd}
We simulate the momentum-resolved reflectivity of the 2D crystal and -- as a reference -- an empty space in the finite-difference time-domain (FDTD) using the package MEEP~\cite{oskooi2010elsevier}.
An $s$- or $p$-polarized Gaussian pulse with a frequency centered around $\tilde{\nu}_m =$ \pcm{7500} with full width at half maximum $\Delta\tilde{\nu}_m =$ \pcm{5000} is incident on a 2D sample.
The fields are recorded at a distance \um{0.25} from the surface and Fourier transformed in time and space to obtain the momentum-resolved amplitudes $\mathcal{F}\{{\it{A}_{\rm{sample}}}\}$ and $\mathcal{F}\{{\it{A}_{\rm{empty}}}\}$.
The reflectivity as a function of wavenumber $\tilde{\nu}$ and off-axis wave vector $k_y$ is calculated via~\cite{meepTutorial2}
\begin{equation}
    R_{\rm{sample}}(\tilde{\nu}, k_y) = \pct{100}~\frac{\abs{\mathcal{F}\{{\it{A}_{\rm{sample}}}\} - \mathcal{F}\{{\it{A}_{\rm{empty}}}\}}^2}{\abs{\mathcal{F}\{{\it{A}_{\rm{empty}}}\}}^2}.
\end{equation}
The crystal has a size of \numproduct{20 x 6} rectangular unit cells, 
corresponding to 13.6$\,\times\,$ \SI{2.88}{\micro\meter^2}.
To increase the resolution and source size, we pad the structure with \um{10} free space in both $\pm y$ directions.
The maximum $\abs{k_y}$ that we can simulate is then \pum{4.5} using the default GaussianBeam2DSource from MEEP.

\section{Results}\label{sec:pc_res_c_experiment}
\begin{figure}[t]
    \centering
    \includegraphics[width=0.75\linewidth]{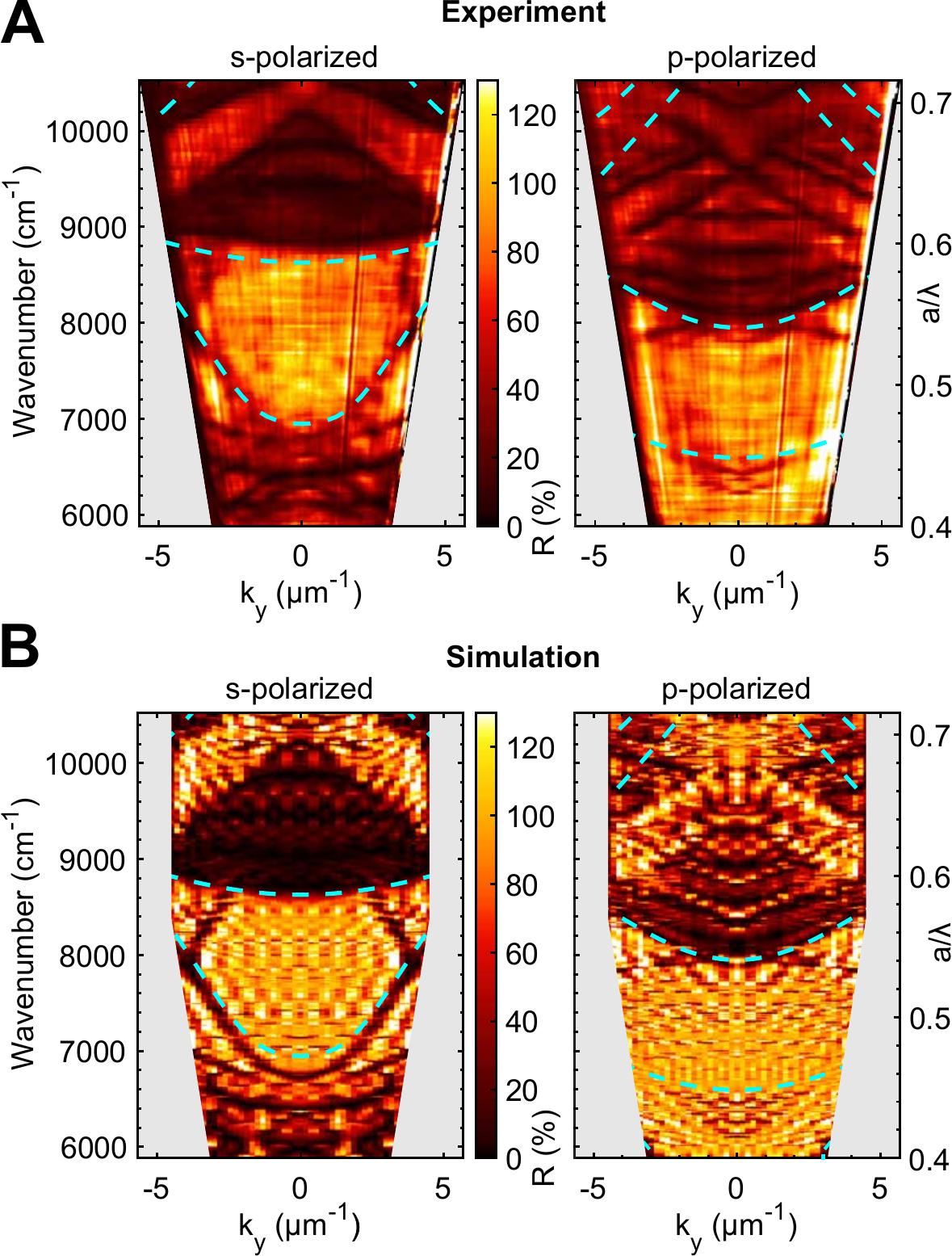}
    \caption{Momentum-resolved reflectivity $R$ of the 2D photonic crystal, where $k_y \equiv k^{\rm{refl}}_y$.
    (Cyan dashed) Predominantly $s$-polarized (left) and $p$-polarized (right) bands calculated on the 11 Bragg plane.
    (a) Experimental data. (b) Simulated data for the same structure.
    }
    \label{fig:pc_res_T2_D}
\end{figure}
\subsection{Experiments \& band calculations}\label{sec:pc_res_exp}
Momentum-resolved reflectivity spectra measured on the 2D crystal are shown in Fig.~\ref{fig:pc_res_T2_D}(a). 
We observe clear features such as deep troughs and high plateaus in the spectra ranging from about 10 to \pct{100} percent reflectivity.
The features are sharp in wavenumber $\tilde{\nu}$, going from high to low reflectivity in a few tens of \pcm{}.
In addition, we observe that the reflectivity is symmetric about $k_y = 0$, which is expected given the symmetry of the sample, the excellent etch quality, and the precisely cut interface plane.

To interpret the features in the reflectivity, we calculate bands (cyan-dashed in Fig.~\ref{fig:pc_res_T2_D}(a)) of this photonic crystal for $\vb{k}$-vectors on the 11 Bragg plane.
The bands are selected on symmetry to be excitable with $s$- or $p$-polarized plane waves.
The sharpest features in the measured reflectivity agree well with the calculated bands:
Firstly, the $s$-polarized band at \pcm{8700} and the $p$-polarized band at \pcm{8100} indicate the edge of a stop gap, with high reflectivity below and low reflectivity above this band.
Secondly, the $s$-polarized band starting at \pcm{7000} and the $p$-polarized band at \pcm{6800} at $k_y = 0$ indicate troughs inside the high reflectivity where the incident waves excite Bloch modes and therefore do not reflect.
The latter band only shows a shallow trough that disappears quickly with increasing $\abs{k_y}$; potentially, the symmetry changes when moving off the axis.

Above \pcm{9000}, the bands calculated on the 11 Bragg plane disagree with the measured reflectivity.
In addition, the bands do not explain many other features, such as the $s$-polarized troughs below \pcm{6800} or the $p$-polarized troughs above \pcm{8500}.
It is possible to further expand the theoretical search and selection of bands by considering bands calculated at other $\kk{}$-vectors, for example, on other Bragg planes; and by considering equifrequency surfaces and the group velocity of Bloch modes to check if particular bands cannot be excited.
However, overall, the bands calculated using 2D theory agree very well with the experiments, showing that the measurements probe 2D nanophotonics.


\subsection{Simulations}\label{sec:pc_res_sims}
To further prove that we measure only in-plane reflectivity, we simulate momentum-resolved reflectivity spectra using 2D FDTD in Fig.~\ref{fig:pc_res_T2_D}(b).
The simulated spectra exhibit sharp features with reflectivity values ranging from near 0 to \pct{120} and are symmetric about $k_y = 0$.

We again compare the results to the symmetry-selected bands (cyan dashed) calculated at the 11 Bragg plane.
The bands, which are calculated for an infinite photonic crystal, agree well with troughs inside the simulated reflectivity of the finite structure, but -- similar to the experiment -- only for wavenumbers below \pcm{9000}.
Minor discrepancies occur because the band calculations consider an infinite structure while the simulations consider a finite width, $L = 6c$.

The simulated 2D reflectivity of Fig.~\ref{fig:pc_res_T2_D}(b) agrees very well with the measured effectively 2D experiment of Fig.~\ref{fig:pc_res_T2_D}(a).
Around the calculated bands below \pcm{9000}, the simulations and experiments show the same troughs and steps in reflectivity.
Besides the reflectivity around the bands, for $s$-polarized light, experiment and simulation both show upward curving troughs around \pcm{6500}, and a low-reflectivity triangle around \pcm{9500}.
In addition, $p$-polarized features such as the X-shape near \pcm{9700} with the horizontal troughs below and the small troughs inside the high reflectivity at \pcm{8000} match as well.

\begin{figure}
    \centering
    \includegraphics[width=0.7\linewidth]{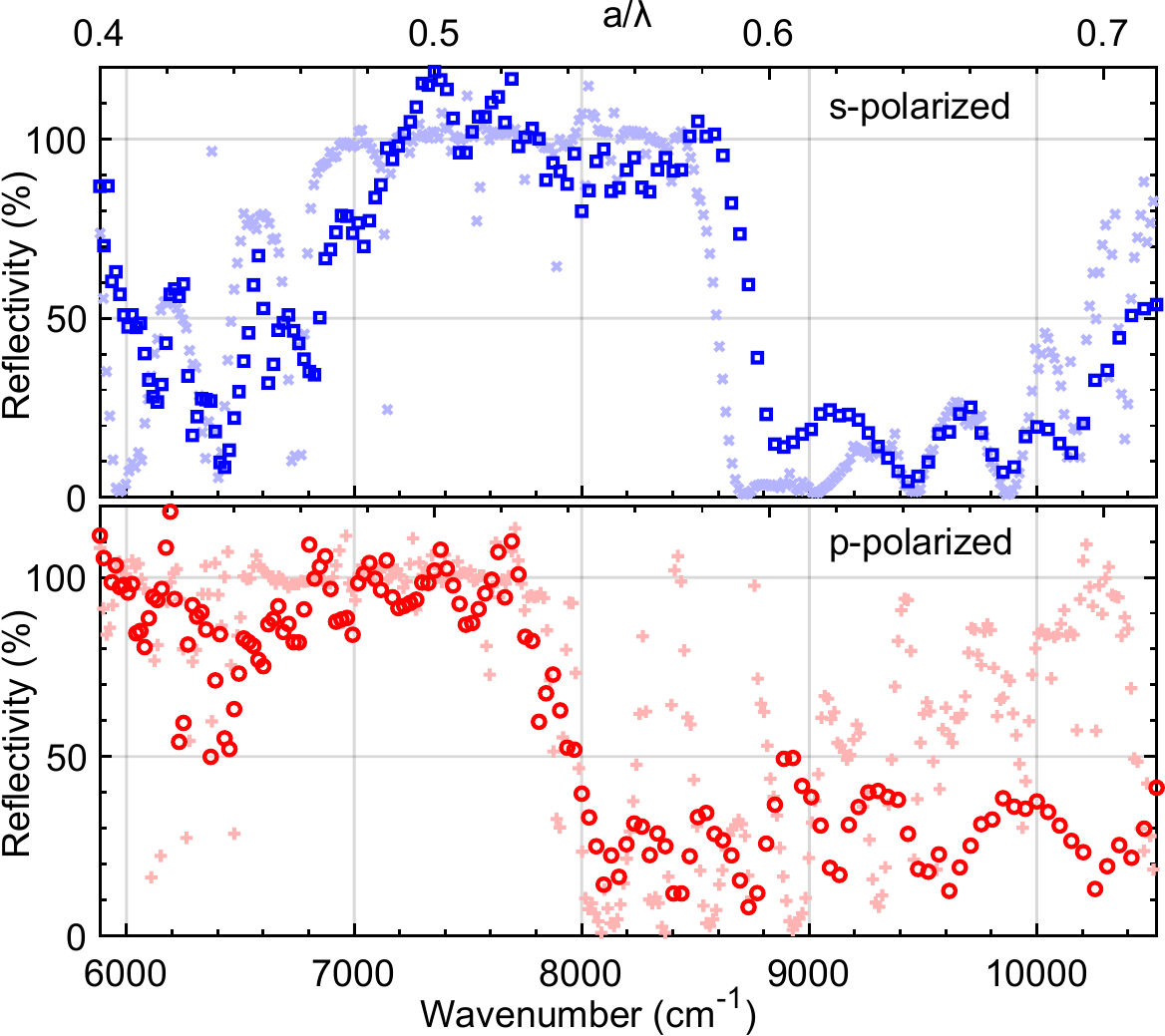}
    \caption{Cross section through the momentum-resolved reflectivity around $k_y = 0$ averaged over wavevectors $k_y/\abs{\vb{k}} \le 0.1$.
    Blue: Experiment $s$-polarized. 
    Light blue: Simulation $s$-polarized experiment.     
    Red: Experiment $p$-polarized.
    Pink: Simulation $p$-polarized.}
    \label{fig:cry2D_crossSec_ky0}
\end{figure}

To quantitatively compare the simulations to the experiments in more detail, we take a cross section through the momentum-resolved spectra around $k_y = 0$ averaged over wavevectors $k_y/\abs{\vb{k}} \le 0.1$ in Fig.~\ref{fig:cry2D_crossSec_ky0}.
We observe that the reflectivities obtained in the experiments and simulations both steeply decrease from a high to a low plateau around $\tilde{\nu} =$ \pcm{7900} for $p$-polarized light and around $\tilde{\nu} =$ \pcm{8600} for $s$-polarized light.
The simulated reflectivity has deeper troughs than in the experiment: the first troughs in the low plateaus reach less than \pct{1} reflectivity in the simulations but \pct{14} in the experiments.
The simulated features are more extreme likely because no manufacturing defects are considered in the simulations.
Specifically, due to manufacturing defects such as splinters in the pores, a varying pore diameter with $z$, and surface roughness, the features soften and smooth out.
We refer to Ref.~\cite{corbijn2023OptExp} for a thorough discussion on the impact of fabrication defects on the reflectivity of 2D photonic crystals.

In the momentum-resolved reflectivity spectra of Fig.~\ref{fig:pc_res_T2_D}(b), we also observe variations in the simulated reflectivity at high spatial frequencies, for example, for $p$-polarized reflectivity at $\tilde{\nu} <$ \pcm{8000}, while the reflectivity is much smoother in the experiment.
We attribute the variations to the effects of the Fourier transform over a finite domain and an imperfect light source in the simulation.
While the precise reflectivity values of the experiments and simulations slightly vary due to fabrication defects, overall the features that we observe agree very well, again confirming that our experimental methods are rigorous.

\section{Conclusion}\label{sec:pc_conclusion}
We have successfully measured the momentum-resolved reflectivity of a 2D photonic crystal in its plane of periodicity in the near-infrared regime.
The reflectivity corresponds excellently to 2D band calculations and 2D momentum-resolved finite-difference time-domain simulations, proving that the experiments only probe wave vectors in the 2D periodic plane.
The results are exciting because photonic crystals and band structures are often studied theoretically in 2D, but to the best of our knowledge, we are the first to probe them experimentally in the near-infrared spectral regime around $\lambda =$ \nm{1550}.
Therefore, we believe our results can be used to further understand complex phenomena in the field of photonic crystals, such as the roles of symmetry and group velocity of the incident and excited waves in the percentage of light that is reflected or transmitted.
The optical procedure is readily applicable to other 2D structures, such as 2D quasicrystals, as well as 3D structures, including 3D photonic crystals, and structures with intentional and unintentional defects.

\begin{backmatter}
\bmsection{Funding}
This work was supported by the project ``Self-Assembled Icosahedral Photonic Quasicrystals with a Band Gap for Visible Light'' (OCENW.GROOT.2019.071) of the ``Nederlandse Organisatie voor Wetenschappelijk Onderzoek'' (NWO); 
the NWO-TTW Perspectief program P15-36 ``Free-form scattering optics'' (FFSO) with TUE, TUD, ASML, Demcon, Lumileds, Schott, Signify, and TNO; 
and by the NWO-TTW Perspectief program P21-20 ``Optical coherence; optimal delivery and positioning'' (OPTIC) with TUE, TUD, ARCNL, Anteryon, ASML, Demcon, JMO, Signify, and TNO.

\bmsection{Acknowledgment}
We thank Lars Corbijn van Willenswaard and Thomas Krauss for helpful discussions, and Manashee Adhikary and Geert-Jan Kamphuis for experimental help.
We thank the MEEP and MPB teams for their extensive examples.

\bmsection{Disclosures}
The authors declare no conflicts of interest.

\bmsection{Data availability}
Data underlying the results presented in this paper are available when published.

\end{backmatter}

\bibliography{references_2D-phc} 

\end{document}